\documentclass[acmlarge]{acmart}
\usepackage{graphicx}
\usepackage{subfigure}
 \usepackage{algorithm}
\usepackage{algorithmic}
\usepackage{booktabs} 
\usepackage{amsmath}
 \usepackage{multirow}
 \usepackage{rotating}
 
 \usepackage{amssymb}
 \usepackage{bbding}
 \usepackage{amsthm}
 \usepackage{bm}
 \usepackage{epstopdf}
 \usepackage{enumerate}
 \usepackage{array}
\usepackage{bbm}

\AtBeginDocument{%
  \providecommand\BibTeX{{%
    \normalfont B\kern-0.5em{\scshape i\kern-0.25em b}\kern-0.8em\TeX}}}

\setcopyright{acmcopyright}
\copyrightyear{2020}
\acmYear{2020}
\acmDOI{10.1145/1122445.1122456}

\acmJournal{TOIS}
\acmVolume{37}
\acmNumber{4}
\acmArticle{111}
\acmMonth{8}

\begin{document}

\title{CoSam: An Efficient Collaborative Adaptive Sampler for Recommendation}

\author{Jiawei Chen}
\email{sleepyhunt@zju.edu.cn}
\author{Chengquan Jiang}
\email{	imjcqt@gmail.com}
\author{Can Wang$^\dagger$}
\thanks{$^\dagger$Corresponding author: wcan@zju.edu.cn}
\email{wcan@zju.edu.cn}
\author{Sheng Zhou}
\email{	zhousheng_zju@zju.edu.cn}
\author{Yan Feng}
\email{fengyan@zju.edu.cn}
\author{Chun Chen}
\email{chenc@cs.zju.edu.cn}

\affiliation{%
  \institution{Zhejiang University}
  \streetaddress{38 ZheDa Road}
  \city{Hangzhou}
  \country{China}
}

\author{Martin Ester}
\affiliation{
  \institution{Simon Fraser University}
  \streetaddress{8888 University Drive}
  \city{Burnaby}
  \country{Canada}}
\email{ester@cs.sfu.ca}

\author{Xiangnan He}
\affiliation{%
  \institution{University of Science and Technology of China}
  \streetaddress{96 JinZhai Road}
  \city{Hefei}
  \country{China}
}
\email{xiangnanhe@gmail.com}

\renewcommand{\shortauthors}{Chen and Jiang, et al.}

\begin{abstract}
 Sampling strategies have been widely applied in many recommendation systems to accelerate model learning from implicit feedback data. A typical strategy is to draw negative instances with uniform distribution, which however will severely affect model's convergency, stability, and even recommendation accuracy. A promising solution for this problem is to over-sample the ``difficult'' (a.k.a informative) instances that contribute more on training. But this will increase the risk of biasing the model and leading to non-optimal results. Moreover, existing samplers are either heuristic, which require domain knowledge and often fail to capture real ``difficult'' instances; or rely on a sampler model that suffers from low efficiency.

To deal with these problems, we propose an efficient and effective collaborative sampling method CoSam, which consists of: (1) a collaborative sampler model that explicitly leverages user-item interaction information in sampling probability and exhibits good properties of normalization, adaption, interaction information awareness, and sampling efficiency; and (2) an integrated sampler-recommender framework, leveraging the sampler model in prediction to offset the bias caused by uneven sampling. Correspondingly, we derive a fast reinforced training algorithm of our framework to boost the sampler performance and sampler-recommender collaboration. Extensive experiments on four real-world datasets demonstrate the superiority of the proposed collaborative sampler model and integrated sampler-recommender framework.
\end{abstract}



\maketitle

\section{Introduction}
\allowdisplaybreaks[4]

\begin{figure}[t]
  \centering
  \includegraphics[width=0.7\textwidth]{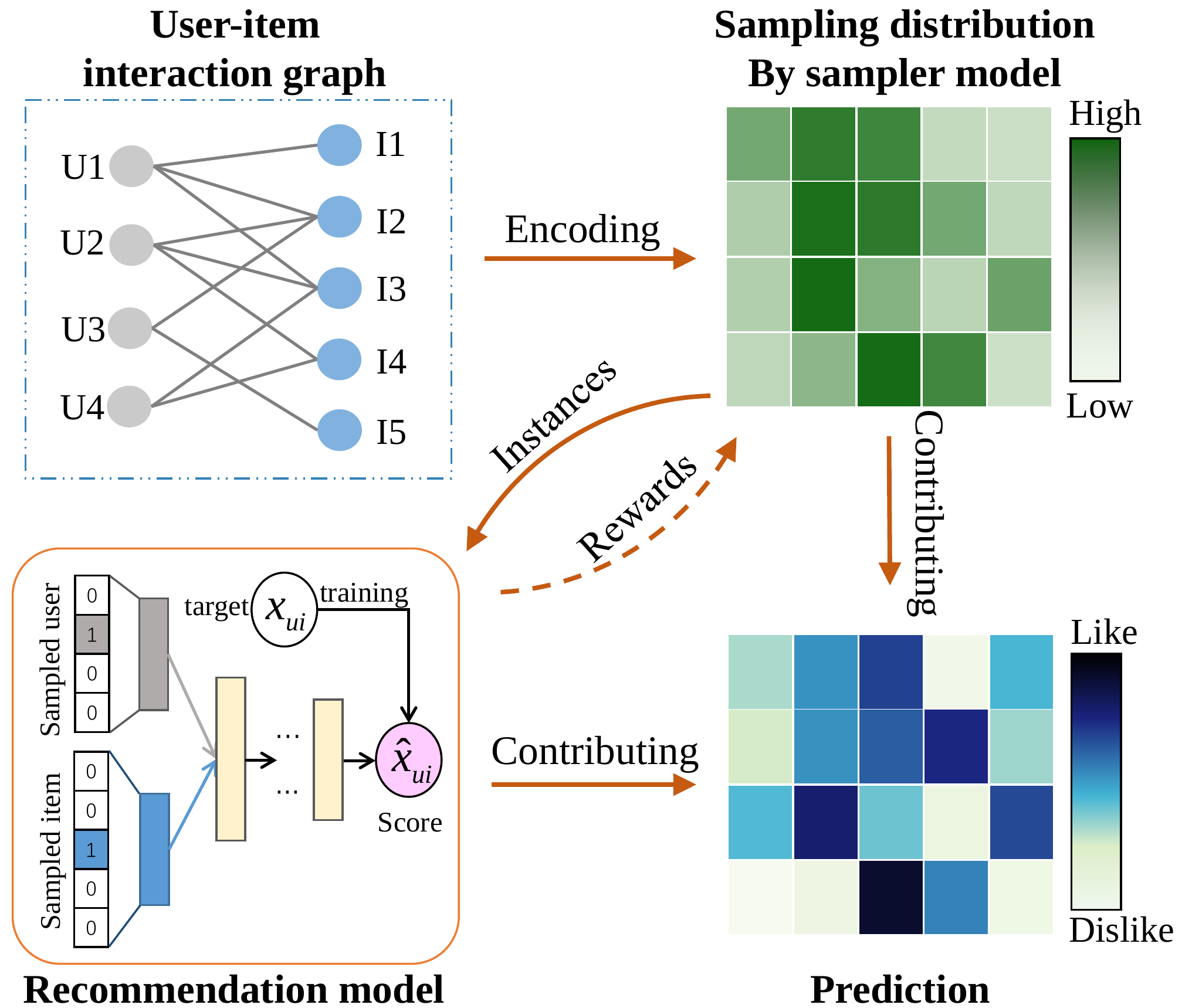}\\
  \caption{Illustration of CoSam, which leverages interaction graph into sampling and integrates sampling distribution (model) into prediction. The user-item interaction graph is constructed from user-item interaction information, where we give links for the user-item pair with positive feedback. }\label{fg:shiyi}
   \vspace{-0.4cm}
\end{figure}

Recent research attention on recommendation is increasingly shifting from explicit feedback data (a.k.a. rating data) towards implicit feedback data. Implicit feedback data is often a natural byproduct of users' behavior, and thus is more prevalent, inexpensive and scalable. However, recommendation from implicit feedback is very challenging due to the lack of negative feedback data. While the unobserved feedback data can be treated as a source of negative signal, its large scale incurs expensive time cost in learning a recommendation model.  Existing methods mainly rely on stochastic gradient descent optimizer (SGD) and negative sampling strategies to speed up training procedures. Since the gradients are estimated on the sampled instances, sampling strategies play an important role in recommendation systems and significantly affect the training efficiency and recommendation accuracy.

While there are rich literatures on sampling strategies for better recommendation, to our knowledge, existing samplers lack one or more desiderata. They can be roughly classified into the following four types:
\begin{itemize}
\item \textbf{Uniform sampler,} the most popular strategy that samples negative instances with equal probability. Although efficient, it usually causes slow convergence and high variance, leading to non-optimal results  \cite{rendle2014improving,yu2018walkranker,yu2017selection}.
\item \textbf{Heuristic-based samplers,} which define specific sampling rules to over-sample the ``difficult'' (a.k.a informative) negative instances. The intuition behind this type of samplers is that the ``difficult'' instances, which can be defined as the items with higher popularity \cite{yu2017selection} or higher ranking position \cite{rendle2014improving}, make more contribution on gradients. However, these methods involve heuristic alterations to the model and data, requiring recommendation expertise or tedious hyper-parameter tuning. Moreover, heuristic samplers usually lack flexibility and often fail to capture really ``difficult'' instances.
\item \textbf{Model-based adaptive samplers,} which define a class of sampling distribution with a flexible sampler model \cite{wang2017irgan,park2019adversarial,ding2019reinforced}. Existing model-based samplers mainly adopt adversarial learning framework. With adversarial training between the sampler model and the recommender model, the sampler will learn to fit users' preference distribution and thus adaptively generate positive-like ``difficult'' instances. While this type of samplers have been proved effective, they tend to be very slow in large dataset. The reason is that the sampling distribution will dynamic evolve with the training proceeding, the sampling process usually requires a traversal of all instances, which is time-consuming and computationally inefficient. Moreover, the adversarial sampler will yield a biased estimate of the gradients since its uneven sampling distribution will skew the instance contribution, resulting in sub-optimal results. Typically, the recommendation model will be biased to over-fit the ``difficult'' negative instances. These instances will be predicted with relatively low preference scores. In fact, these ``difficult'' but positive-like items are likely to be adopted by the user.
\item \textbf{Auxiliary information-based samplers,} which leverage auxiliary information (e.g. social network, users' exposure) to sample informative negative instances \cite{chen2019samwalker,ding2019reinforced}. However, these auxiliary information may not be available in many recommender systems.
\end{itemize}

In a word, designing a satisfactory sampler is important, but challenging. Drawing on the advantages of existing model-based samplers, we explore a more advanced model-based sampler. We have to address the following key research problems:
\begin{enumerate}[(P1)]
\item How to design a sampler model that supports efficient and effective sampling?
\item How to mitigate the bias caused by the uneven sampling probability?
\end{enumerate}
In this paper, we propose a novel model-based adaptive sampler named \textbf{Co}llaborative \textbf{Sam}pler (CoSam). Corresponding to two problems, CoSam has two improvements and ``collaborative'' has two-fold meanings. First, to tackle the problem (P1), we propose a novel \emph{collaborative} sampler model, which leverages \emph{collaborative} signals in sampling. The collaborative signals --- more specifically the user-item interaction information and its extended user-item interaction graph as illustrated in Figure \ref{fg:shiyi}--- reveals correlations among users and items. These correlations can be expressed with the information paths along the graph. For example, as illustrated in Figure \ref{fg:shiyi}, the information paths U1-I2-U2, U1-I3-U2 between users U1 and U2, indicate the users have similar behaviors, as they have interacted with common items I2, I3. Correspondingly, their feedback data may have similar level of difficulty and should be sampled with similar distribution. Another example can be seen from the rich information paths between the user U1 and item I4 (e.g. U1-I2-U2-I4, U1-I3-U4-I4). It reflects U1 and I4 may have high affinity although they have no directed interaction, since U1's similar users U2,U4 have interacted with I4. Correspondingly, this feedback data is potentially highly informative (difficult) for the recommendation model. Thus, CoSam defines a sampler model on the graph and constructs adaptive sampling distribution based on information paths, which naturally encodes this useful correlation information into distribution and potentially boosts sampling performance. Moreover, a specific random walk-based sampling strategy is developed to support fast sampling from the proposed dynamic distribution, which will evolve over the large item space as training goes on. Our theoretical analysis proves that the CoSam satisfies the desirable properties of normalization, adaption, interaction information awareness, and sampling efficiency.

Second, to address the problem (P2), we propose a novel integrated sampler-recommender framework to replace adversarial framework. In our integrated framework, the sampler acts as a filter to remove relatively ``easy'' instances and the recommender focuses on differentiating the really ``difficult'' instances that can not be differentiated by the sampler. This way, the sampler and the recommender will \emph{collaboratively} characterize users' preference and satisfy the following desirable property: even although the recommender model is learned with an highly uneven sampler, its biased prediction can be refined by integrating the sampling probability. We further derive a fast reinforced training algorithm of our framework to boost sampler's performance and sampler-recommender \emph{collaboration}. Our framework and algorithm can be used for most recommender models.

It is worthwhile to highlight the following contributions:
\begin{itemize}
\item We propose an efficient and effective collaborative sampler model by leveraging user-item interaction information into sampling.
\item We propose an integrated sampler-recommender framework, which leverages the sampler model into prediction to offset the sampling bias.
\item Our experimental evaluations on four well-known benchmark datasets validate the superiority of the proposed collaborative sampler model and integrated sampler-recommender framework.
\end{itemize}

The rest of this paper is organized as follows. We give the problem definition and background in section 2. The proposed sampling method CoSam is introduced in section 3. The experimental results and discussions are presented in section 4. We briefly review related works in section 5. Finally, we conclude the paper and present some directions for future work in section 6.

\section{Preliminaries}
In this section, we first give the problem definition of the implicit recommendation. Then, we introduce the model-based sampler with the adversarial sampler-recommender framework and analyze its merits and weaknesses.
\subsection{Recommendation from implicit feedback}
Suppose we have a recommender system with user set $U$ (including $n$ users) and item set $I$ (including $m$ items). The implicit feedback data is represented as a $n \times m$ binary matrix $\mathbf X$ with entries $x_{ui} \in \{0,1\}$ denoting whether or not the user $u$ has interacted the item $i$. Also, we let $\mathcal X_u$ denote the set of positive items that have been interacted by user $u$. The task of an implicit recommender system can be stated as follow: learning user's preference from the available implicit feedback and recommending items that are most likely to be interacted by the user.
\subsection{Adversarial sampler-recommender framework}
The adversarial sampler-recommender framework contains:

\textbf{a sampler model (S)}: $G_u\sim Multinomial(N_u,p_s(i|u;\theta))$, which draws a small item set ($G_u$) for each user $u$ from the global item space; where the parameter $N_u$ denotes the number of sampled items and the conditional probability ${p_s}(i|u;\theta)$ denotes the flexible sampling distribution with parameters $\theta$. The sampler model tries to fit the user's preference distribution to generate ``difficult'' instances for recommender. We remark that some works \cite{ding2019reinforced} will remove the sampled positive instances from $G_u$.

\textbf{a recommender model (R)}: which captures user's preference and makes recommendation. In the adversarial framework, R is optimized to distill real positive items ($i\in \mathcal X_u$) from the sampled training set $[\mathcal X_u,G_u]$: $\mathcal X_u\sim p_r(\mathcal X_u|[\mathcal X_u,G_u],f_r(u,i;\varphi))$, where $[\mathcal X_u,G_u]$ denotes the union of the positive item set $X_u$ and sampled negative item set $G_u$. $f_r(u,i;\varphi)$ is a parameterized preference function, mapping attributes (e.g. ids) of users and items into the predicted preference with parameters $\varphi$; $p_r$ denotes likelihood function. Without loss of generality, here we choose Bernoulli classifier as an example for further derivation and it can be easily replaced by other functions (e.g. Gaussian likelihood \cite{Pan2008}, Bayesian Personalized Ranking \cite{rendle2009bpr}). Formally, we have:
\begin{align}
p_r(\mathcal X_u|[\mathcal X_u,G_u]) = \prod\limits_{i \in {[\mathcal X_u,G_u]}} {Bernoulli({x_{ui}},f_r(u,i;\varphi))}
\end{align}

In the adversarial framework, S and R play an adversarial game and optimize the following objective function:
\begin{align}
\mathop {\min }\limits_\theta  \mathop {\max }\limits_\varphi  \sum\limits_{u \in U} {{E_{{G_u} \sim S}}[\sum\limits_{i \in {X_u}} {\log ({f_r}(u,i;\varphi ))}  + \sum\limits_{i \in {G_u}} {\log (1 - {f_r}(u,i;\varphi ))} ]}
\end{align}
where $E_{G_u\sim S}[.]$ stands for $E_{G_u\sim Multinomial(N_u,p_s(i|u;\theta))}[.]$. Naturally, with the competition between two models, S will be learned to draw the difficult instances with higher preference scores, which provide larger gradients and more information to train R.

\textbf{Weaknesses.} However, although the model-based sampler with adversarial sampler-recommender framework is capable of adaptively capturing difficult instances, we emphasize two weaknesses of this kind of methods: (1) The sampling process is time-cost expensive. Since the sampling probability will evolve over the large item space with the training going on, the sampling process requires a traversal of all instances ($(O(n\times m))$) to update the dynamic sampling distribution and to draw instances from the distribution. (2) The adversarial uneven sampler will create biases on the recommendation model, leading to non-optimal results. It can be understood from the expected objective function of the recommender model in the adversarial framework:
\begin{align}
\sum\limits_{u \in U} {[\sum\limits_{i \in{\mathcal X_u}} {\log ({f_r}(u,i;\varphi ))}  + \sum\limits_{i \in I\backslash {\mathcal X_u}} {N_u^rE[{p_r(i|u)}]\log (1 - {f_r}(u,i;\varphi ))} ]}
\end{align}
which is different from the original objective function without sampling. The negative instances have been weighted with sampling probability and the objective function is biased from the data likelihood. Moreover, with the adversarial training between the sampler and recommender, the sampler will highly incline to draw "positive-like" negative data \cite{goodfellow2014generative,gao2018self}. Correspondingly, these instances will be over-trained and the learned preference scores ${f_r}(u,i;\varphi )$ of these instances will be biased to have relatively low values. In fact, these negative but ``positive-like'' items are highly likely to be interacted by the user. The recommendation performance will suffer. Meanwhile, the prediction accuracy on other relatively ``easy'' instances may deteriorate too due to the insufficient training of them. A naive solution for this problem is to offset the training instances with the inverse of their sampling probability. However, the gradients will be highly instable and potentially exploded.

Thus, in this paper,  we are interested in, and propose a solution to two related problems:
\begin{enumerate}
\item A new sampler model support efficient and effective sampling.
\item A new sampler-recommender framework to mitigate biases caused by uneven sampler.
\end{enumerate}

\section{Methodology}
We now present the proposed sampler CoSam. there are two key components in CoSam: a collaborative sampler model and an integrated sampler-recommender framework.

\subsection{Collaborative sampler model}
\label{se:random}
To support both effective and efficient sampling, we propose a new sampler model based on the interaction graph structure, which contains rich correlations information among users and items. Inspired by the success of the graph convolutional network model \cite{wang2019neural,ying2018graph,kipf2016semi} in capturing graph proximity, CoSam contains many propagation layers, which refine users sampling probability based on their graph neighbors. Let vector $\bm \rho_u$ denote the sampling distribution for each user $u$ over items ($p_s(i|u;\varphi)\equiv \bm \rho_u$). We initially set $\bm \rho^{(0)}_u$ with uniform distribution $\mathcal U$ and set the initial label for each item with its item indicator vector $\bm \gamma^{(0)}_{i}=\mathbf z_i$. $\mathbf z_i$ is an one-hot vector in which $i$-th element is equal to 1. The model will propagation users and items information along the graph to refine users' sampling distribution and naturally inject useful correlation information into sampling distribution. Concretely, as shown in Figure \ref{fg:gnn}, in each layer, each node in the graph collect information from the connected neighbors and reconstruct their distribution or labels as follows:
\begin{align}
\bm \rho _u^{(t)} = c_1\sum\limits_{i \in I,i \in {N_u}} {w_{ui}^{(1)}} \bm \gamma _i^{(t-1)} + (1-c_1)\bm \rho _u^{(0)} \label{eq:it1} \\
\bm \gamma _i^{(t)} =c_2\sum\limits_{u \in U,u \in {N_i}} {w_{iu}^{(2)}\bm \rho _u^{(t-1)}}  + (1-c_2)\bm \gamma _i^{(0)} \label{eq:it2}
\end{align}
The parameters $c_1,c_2$ $(0 \le  c_1,c_2 \le  1)$ specify the relative contributions from the neighbors and the initial vectors. $w_{ui}^{(1)},w_{iu}^{(2)}$ are learned parameters, which balances the heterogenous influence from different neighborhoods and meets $\sum\limits_{i \in I,i \in {N_i}} {w_{ui}^{(1)}}  = 1$, $\sum\limits_{u \in U,u \in {N_i}} {w_{iu}^{(2)}}  = 1, w_{ui}^{(1)}>0, w_{ui}^{(2)}>0$.
Overall, CoSam models sampling probability with an interactive graph-aware function $g$ parameterized by edge weights $w$ ($w^{(1)},w^{(2)}$), to which eq. (\ref{eq:it1}),(\ref{eq:it2}) converges:
\begin{align}
\begin{bmatrix}
\mathbf P\\
\mathbf Y
\end{bmatrix} = {g_w}(\mathbf X) \equiv \mathop {\lim }\limits_{t \to \infty } \begin{bmatrix}
{{\mathbf P^{(t)}}}\\
{{\mathbf Y^{(t)}}}
\end{bmatrix} = {(\mathbf I - \mathbf C{\begin{bmatrix}
\mathbf 0&{{\mathbf W^{(1)}}}\\
{{\mathbf W^{(2)}}}&\mathbf 0
\end{bmatrix}} )^{ - 1}}(\mathbf I-\mathbf C)\begin{bmatrix}
{{\mathbf P^{(0)}}}\\
{{\mathbf Y^{(0)}}}
\end{bmatrix} \label{eq:co}
\end{align}
where we collect sampling distribution $\bm \rho_u$ for each user as matrix $\mathbf P$ and collect labels $\bm \gamma_i$ for each item as matrix $\mathbf Y$; collect edge weights $w^{(1)}_{ui}, w^{(2)}_{iu}$ as matrixes $\mathbf W^{(1)}, \mathbf W^{(2)}$, in which $\mathbf W^{(1)}_{ui}=w^{(1)}_{ui}$, $\mathbf W^{(2)}_{iu}=w^{(2)}_{iu}$ for connected user-item pairs and $\mathbf W^{(1)}_{ui}=0, \mathbf W^{(2)}_{iu}=0$ for others; collect $c_1,c_2$ as a $(n+m)\times (n+m)$ diagonal matrix $\mathbf C$, in which the first $n$ diagonal elements are $c_1$ and other diagonal elements are $c_2$. We show that the proposed collaborative sampler model satisfies the following four desirable properties:

\begin{figure*}[t!]
  \centering
  \includegraphics[width=0.8\textwidth]{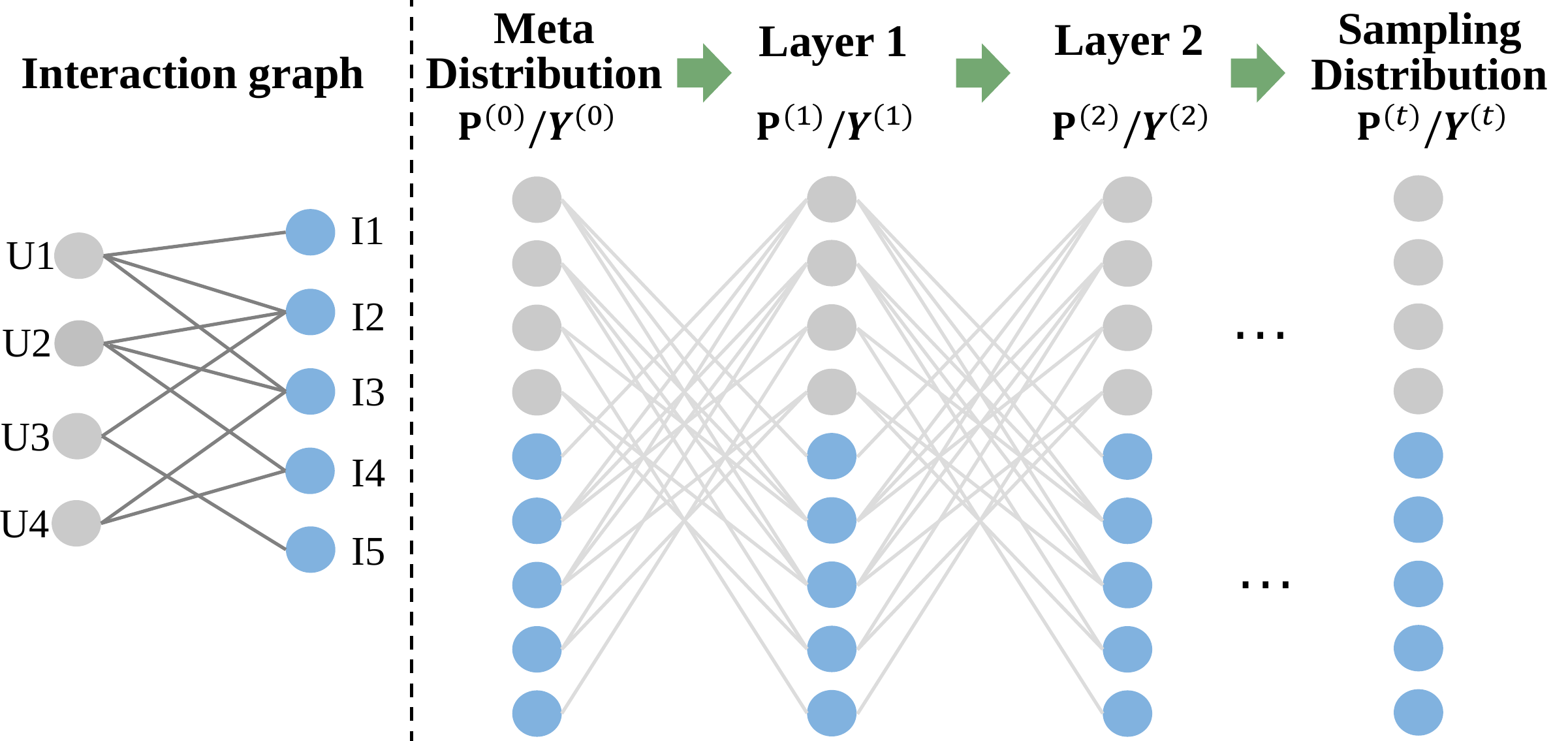}\\
  \caption{Illustration of the collaborative sampler model.}\label{fg:gnn}
\end{figure*}

\begin{figure*}[t!]
  \centering
  \includegraphics[width=0.95\textwidth]{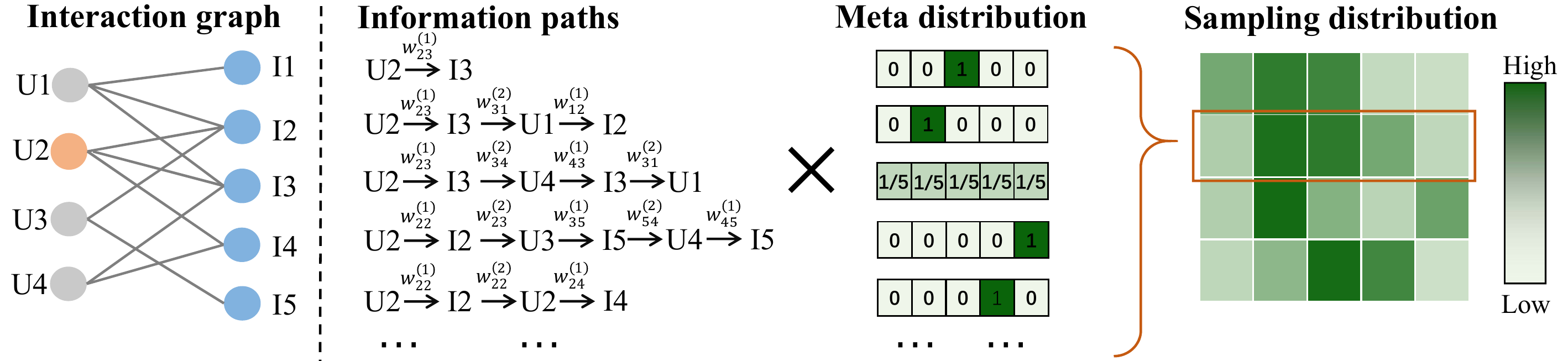}\\
  \caption{The constructed sampling distribution of CoSam can be considered as a mixture of meta distributions, where the mixing coefficients are controlled by the information paths and the learned path strength.}\label{fg:shisam}
\end{figure*}

\textbf{Normalized.} The constructed sampling model must have a valid probability distribution over the item space, i.e. for each user $u$, we have $\sum\limits_{i \in I} {{\bm \rho _{ui}}}  = 1$.
\begin{proof}
Let mark the adjacency matrix of the interactive graph as $\mathbf W=\begin{bmatrix}
\mathbf 0&{{\mathbf W^{(1)}}}\\
{{\mathbf W^{(2)}}}&\mathbf 0
\end{bmatrix}$. Here we mark user nodes in the interaction graph with id $1$ to $n$, and item nodes with id $n+1$ to $n+m$. Let $\mathbf e$ denotes the column vector of $n+m$ ones. we have: $(I-\mathbf C\mathbf W)\mathbf e=(I-\mathbf C)\mathbf e$. Note that $(I-\mathbf C\mathbf W)^{-1}(I-\mathbf C\mathbf W)\mathbf e=\mathbf e$. Thus, we can get $(I-c\mathbf W)^{-1}(I-\mathbf C)\mathbf e=\mathbf e$. Then, we have $\begin{bmatrix}
{{\mathbf P^{(t)}}}\\
{{\mathbf Y^{(t)}}}
\end{bmatrix}\mathbf e=(I-c\mathbf W)^{-1}(I-\mathbf C)\begin{bmatrix}
{{\mathbf P^{(0)}}}\\
{{\mathbf Y^{(0)}}}
\end{bmatrix}\mathbf e=\mathbf e$.
\end{proof}
\textbf{Interaction information awareness.} As we can seem from eq.(\ref{eq:co}), the matrix $\mathbf H=(I-CW)^{-1}(I-C)$ is a graph or diffusion kernel \cite{zhou2004learning}, which has been widely adopted to measure nodes proximity in the graph. Thus, as shown in Figure \ref{fg:shisam}, the constructed sampling distribution can be considered as a mixture of the meta distributions (uniform distribution or item indictor distribution), where the mixing coefficients are specified with the adaptive interaction graph proximity, which will evolve with training process going on. That is, we have the following formula of the sampling distribution of user $u$:
\begin{align}
{\rho _u} = \sum\limits_{k \in U} {{\mathbf H_{u,k}}\mathcal U}  + \sum\limits_{k \in I} {{\mathbf H_{u,n + k}}{\mathbf z_k}}
\end{align}
where $\mathbf H_{i,j}$ denotes the $(i,j)$-th element in the matrix $\mathbf H$ characterizing the graph proximity between the nodes $i$ and $j$. the graph proximity $\mathbf H_{i,j}$ can be further depicted as aggregated strength of different information paths between the nodes $i$ and $j$:
\begin{align}
\mathbf H_{ij}&=((1-c_1) + (1-c_2)c_1{\mathbf W_{i,j}} + (1-c_1)c_1c_2\prod\limits_{{k_1}\in I} {{\mathbf W_{i,{k_1}}}{\mathbf W_{{k_1},j}}} \notag \\ &+ (1-c_2)c_1c_2c_1\prod\limits_{{k_1}\in I,{k_2}\in U} {{\mathbf W_{i,{k_1+m}}}{\mathbf W_{{k_1+m},{k_2}}}{\mathbf W_{{k_2},j}}}+\ldots) \notag \\
&=\sum_{\mathcal P \in \mathcal A_{ij}} F(\mathcal P) \label{eq:hp}
 \end{align}
where $\mathcal P$ denotes a path instance on the graph and $\mathcal A_{ij}$ denotes the set of path $\mathcal P$ between node $i$ and $j$. $F(\mathcal P)$ is the product of edge weights on the path, which can be defined as path strength between the nodes. The path with shorter length and larger edge weights will have stronger path strength. Thus, as illustrated in Figure \ref{fg:shisam}, the constructed sampling probability can be considered as a mixture of meta distributions, where the mixing coefficients are controlled by the number of information paths and adaptive paths strength. A closer user-item pair with more and stronger information paths will be sampled with higher probability. Naturally, the rich interaction information has been encoded into sampling distribution.

\textbf{Sampling efficiency.} Although the sampling distribution of CoSam will evolve with training going on, CoSam supports fast sampling with the following adaptive random walk strategy:

\emph{Adaptive random walk strategy (ARW for short):} For the target user $u$, we perform the adaptive random walk from $u$ to sample the instances from ${\rho _u}$: At each step of random walk, we are at a certain node $v$. We have two options: (1) If $v$ is a user (or item) node, with probability $(1-c_1)$ (or $(1-c_2)$), we terminate the random walk. we randomly sample an item with meta distribution (or select $v$ as an instance). (2) With probability $c_1$ (or $c_2$), we randomly transfer to one of $v$'s neighbors based on current learned weights $W_{v*}$.

It is easy to check that the path strength is equal to the sampled probability of the walking path from the aforementioned adaptive random walk strategy (ARW for short): $F(\mathcal P|\mathcal P \in \mathcal A_{uv})=p_{ARW}(\mathcal P,v|u,\theta)$. Thus, the sampled items with ARW meets distribution $\rho_u$, even though $\rho_u$ will evolve with training going on. This way, the sampling process just requires a random walk along the graph instead of an expensive traversal of all items. the complexity of sampling an instance can be heavily reduced from $O(m)$ to $O(l_r)$, where $l_r$ denotes the the length of the random walk and its expectation is upper-bounded with $E[l_r]=2/(1-min(c_1,c_2))$. In practical, we are usually interested in users' local graph contexts and thus control the length of random walk smaller than 10, which is much smaller than the number of items.

We remark that the training and updating of the collaborative sampler is efficient too, which will be discussed in the subsection \ref{se:coif}.

\textbf{Adaption.} We remark the adaption of our collaborative sampler model in following two aspects: (1) The sampling distribution of CoSam is defined with a parameterized function instead of pre-defined values. Thus, the sampling distribution will adaptively evolve with training process going on. (2) The random walk-based sampling strategy has adaptive transfer probability so that it supports adaptively sampling from dynamic sampling distribution.

\subsection{Integrated sampler-recommender framework}

 Here we present our novel integrated sampler-recommender framework. Instead of letting the sampler (S) and recommender (R) play an adversarial game, we integrate them into a probabilistic generative process. The framework is motivated by the recent business study of the consumer behavior. As suggested in \cite{bray2008consumer}, consumers will first collect a set of potentially-preferred candidate items, and then carefully select desired items from the set. Correspondingly, S and R are integrated into the following decision-making process:
\begin{enumerate}
\item Sample a small potential-preferred item set $G_u$ from the global item set based on S: \\
 $G_u\sim Multinomial(N_u,p_s(i|u;\theta))$.
\item Distill truly positive items from $G_u$ based on R:\\
 $\mathcal X_u\sim p_r(\mathcal X_u|G_u,f_r(u,i;\varphi))$.
\end{enumerate}
The intuition behind this process is similar to the adversarial framework: S acts as a filter to remove relatively ``easy'' instances and thus R can be guided to focus on the really ``difficult'' instances that can not be differentiated by S. But the integrated framework differs from the adversarial framework in that the sampler in our framework will make contribution on prediction. S and R in our framework collaboratively characterize users' preference. Concretely, the probability of the item being positive can be depicted as follow:
\begin{align}
p(i \in {\mathcal X_u}) &= p(i \in {G_u})p(i \in {\mathcal X_u}|i \in {G_u}) \notag\\ &\propto {p_s}(i|u;\theta){f_r }(u,i;\varphi) \label{eq:re1}
\end{align}
 A good property of our framework can be observed: even although the recommender model is trained with an highly uneven sampler, its biased prediction can be refined by integrating the sampling probability. It is coincident with our intuition. R's biased prediction on the ``difficult'' instances, which may be biased with low values due to the over-training, will be refined with multiplied by a relatively large value; On the contrary, the prediction on these ``easy'' instances, which is highly negative, will be weighted with small values. Equation (\ref{eq:re1}) guarantees the final prediction will fit the data likelihood.

\subsubsection{Training objective and algorithm.\\}
We further derive a fast and general training algorithm to learn the sampler model (S) and recommender model (R) in our framework. To make unbiased prediction, S and R are optimized by maximizing the following data likelihood:

\begin{align}
\log p(X) &= \sum\limits_{u\in U} {\log p({\mathcal X_u})}  = \sum\limits_{u\in U} {\log \sum\limits_{{G_u}} {p({\mathcal X_u},{G_u})} }  \notag \\
&= \sum\limits_{u\in U} {\log \sum\limits_{{G_u}} {p({\mathcal X_u}|{G_u})p({G_u})} } \label{eq:marg}
\end{align}

However, directly optimizing marginal probability $p(X)$ is intractable due to the sum over all possible sampled sets inside the logarithm. The number of the feasible item sets is exponential growth, which causes efficiency problem. To deal with this problem, we do some derivations of equation (\ref{eq:marg}) for fast learning.

Note that the positive instances must be contained in the sampled set, otherwise they can not be selected by the model R. Thus, we can divide the $G_u$ into two disjoint parts: the positive item set $G^+_u$ , which is observed constant and equivalent to $\mathcal X_u$; the rest item set $G^r_u$ containing other sampled items. Thus, the probability of the sampled item set can be approximated as: $p(G_u)\approx C_{{N_u}}^{|{X_u}|} p(G_u^+)p(G_u^r)$, where $C_{{N_u}}^{|{X_u}|}$ denotes Combinatorial Numbers. We have:
\begin{align}
\log p(X) &= \sum\limits_{u \in U} {\log \sum\limits_{{G_u}} {p({\mathcal X_u},{G_u})} } \notag \\
 &= \sum\limits_{u \in U} {\log \sum\limits_{G_u^ r } {p(G_u^ + )p(G_u^ r )p({\mathcal X_u}|G_u^ + ,G_u^ r )} }+const \notag \\
 &= \sum\limits_{u \in U} {(\log p(G_u^ + ) + \log {E_{G_u }}[p({\mathcal X_u}|G_u^ + ,G_u^ r )])}+const \label{eq:tot}
\end{align}
where `const' denotes the constant term and can be left out. However, directly optimizing equation (\ref{eq:tot}) is intractable too due to the expectation over the $G_u^ r$ inside the logarithm. Thus, we turn to optimize the following tight lower bound based on Jensen's inequality:
\begin{align}
\log p(X) &\ge \sum\limits_{u \in U} {(\log p(G_u^ + ) + {E_{G_u }}[\log p({\mathcal X_u}|G_u^ + ,G_u^ r )])} \notag \\
 &= \sum\limits_{u \in U} ( \sum\limits_{i \in {\mathcal X_u}} {\log {p_s}(i|u;\theta)}  + \sum\limits_{i \in {\mathcal X_u}} {\log {f_r }(u,i;\varphi)} + {E_{G_u }}[\sum\limits_{i \in G_u } {(1 - {x_{ui}})\log (1 - {f_r }(u,i;\varphi))} ])\notag \\
 &= \sum\limits_{u \in U} ( \sum\limits_{i \in {\mathcal X_u}} {\log {p_s}(i|u;\theta)}  + \sum\limits_{i \in {\mathcal X_u}} {\log {f_r }(u,i;\varphi)}+ {N_u}{E_{{p_s}}}[(1 - {x_{ui}})\log (1 - {f_r }(u,i;\varphi))])\label{eq:loss}
\end{align}

\textbf{Optimizing recommendation model.} By dropping irrelevant terms in eq.(\ref{eq:loss}), we have the following objective function of R:
\begin{align}
{L_\varphi } = \sum\limits_{i \in {\mathcal X_u}} {\log {f_r }(u,i;\varphi)}  + \sum\limits_{\alpha  \in {G_u},G_u \sim S} {(1 - {x_{u\alpha }})\log (1 - {f_r }(u,\alpha;\varphi ))} \label{eq:ups}
\end{align}
where we perform a sampling approximation in which the item $\alpha$ is from the sampled item set $G_u$ returned by S. We remark that the training objective of R in our framework is consistent with adversarial framework. R is trained on the observed positive data and the sampled negative data from S. Our framework can be easily applied for most recommendation models.

\textbf{Optimizing sampler model.} Due to the discrete nature of candidate set $G_u$, we adopt the policy gradient method from reinforced learning to optimize the model S. That is, the optimization of S can be viewed as a reinforcement learning setting, where R gives a reward to action ``selecting sampled items for a user performed according to the policy probability $p_s(i|u)$. We can derive the policy gradient of $L$ w.r.t $\theta$:

\begin{align}
{\nabla _\theta }L &= \sum\limits_{u \in U} {(\sum\limits_{i \in {\mathcal X_u}} {{\nabla _\theta }\log {p_s}(i|u;\theta)}  + {N_u}{E_{p_s}}[{e_{ui}}])} \notag \\
 &= \sum\limits_{u \in U} {(\sum\limits_{i \in {\mathcal X_u}} {{\nabla _\theta }\log {p_s}(i|u;\theta)}  + {N_u}\sum\limits_{i \in I} {{\nabla _\theta }{p_s}(i|u;\theta){e_{ui}}} )} \notag \\
& = \sum\limits_{u \in U} {(\sum\limits_{i \in {\mathcal X_u}} {{\nabla _\theta }\log {p_s}(i|u;\theta)}  + {N_u}{E_{p_s}}[{\nabla _\theta }\log {p_s}(i|u;\theta){e_{ui}}])} \notag\\
 &\approx \sum\limits_{u \in U} {(\sum\limits_{i \in {\mathcal X_u}} {{\nabla _\theta }\log {p_s}(i|u;\theta)}  + \sum\limits_{\alpha  \in {G_u}, G_u\sim S} {{\nabla _\theta }\log {p_s}(\alpha |u){e_{u\alpha}}} )} \label{eq:upr}
\end{align}
where ${e_{ui}} = (1 - {x_{ui}})\log (1 - {f_\varphi }(u,i ))$ denotes the negative reward for the sampled negative items. Also, we perform a sampling approximation in the last step in which the item $\alpha$ is from the sampled item set $G_u$.

\subsection{Collaborative sampler with integrated framework}
\label{se:coif}
We further integrate aforementioned collaborative sampler into our integrated framework. When collaborative sampler meets integrated framework, the collaborative signals can be well exploited. However, the training of S will suffer from inefficiency due to the heavy computation of the sampling probability and gradients. Thus, we further mitigate this efficiency problem by disassembling the sampling probability with different information paths. we can re-write the sampling probability of the instance in our collaborative sampler as:
\begin{align}
p_s(i|u;\theta)&={\rho _{ui}} = \sum\limits_{k \in U} {{\mathbf H_{u,k}}\frac{1}{m}}  + {{\mathbf H_{u,n + i}}} \notag \\
&=\frac{{1-{c_1}}}{{1 - {c_1}{c_2}}}\frac{1}{m} + \sum\limits_{\mathcal P \in {\mathcal A_{ui}}} {F(\mathcal P)} \notag \\
&\equiv p_0+p_{{ARW}}(\mathcal P,i|u;\theta)
\end{align}
where $F(\mathcal P)$ denotes the path strength defined on eq.(\ref{eq:hp}) and can be considered as the sampled probability of the path from the aforementioned adaptive random walk strategy. Thus, we transfer ${\nabla _\theta }L$ (eq.(\ref{eq:upr})) into:

\begin{align}
{\nabla _\theta }L& = \sum\limits_{u \in U} {(\sum\limits_{i \in {\mathcal X}_u } {{\nabla _\theta }\log {p_s}(i|u;\theta )}  + {N_u}\sum\limits_{i \in I} {{\nabla _\theta }{p_s}(i|u;\theta ){e_{ui}}} )} \notag \\
 &= \sum\limits_{u \in U} {(\sum\limits_{i \in {\mathcal X}_u} {\frac{{\sum\limits_{\mathcal P \in {\mathcal A_{ui}}} {{\nabla _\theta }{p_{ARW}}(\mathcal P,i|u;\theta )} }}{{{p_s}(i|u;\theta )}}}} \notag \\ &{+ {N_u}\sum\limits_{i \in I} {\sum\limits_{\mathcal P \in {\mathcal A_{ui}}} {{\nabla _\theta }{p_{ARW}}(\mathcal P,i|u;\theta ){e_{ui}}} } )} \notag \\
 &= \sum\limits_{u \in U} {(\sum\limits_{i \in {\mathcal X}_u} {\frac{{{E_{\mathcal P,j \sim {p_{ARW}}}}[\mathbf I[j = i]{\nabla _\theta }\log {p_{ARW}}(\mathcal P,i|u;\theta )]}}{{p_0+{E_{\mathcal P,j \sim {p_{ARW}}}}[\mathbf I[j = i]]}}}} \notag \\ &{+ {N_u}{E_{\mathcal P,i \sim {p_s}}}[\log {\nabla _\theta }{p_{ARW}}(\mathcal P,i|u;\theta ){e_{ui}}])}
\end{align}
where $\mathbf I[.]$ denotes indicator function. The gradient can be approximately estimated based on the sampled paths for each user ($\mathcal P \in \mathcal S_u$) in a mini-batch:
\begin{align}
{\nabla _\theta }L&\approx \sum\limits_{u \in U} {(\sum\limits_{i \in {\mathcal X}_u} {\frac{{{\sum_{\mathcal P,j \in \mathcal S_u }}[\mathbf I[j = i]{\nabla _\theta }\log {p_{RAW}}(\mathcal P,i|u;\theta )]}}{{N_up_0+{\sum_{\mathcal P,j \in \mathcal S_u }}[\mathbf I[j = i]]}}}} \notag \\ &{+ {\sum_{\mathcal P,j \in \mathcal S_u }}[\log {\nabla _\theta }{p_{RAW}}(\mathcal P,i|u;\theta ){e_{ui}}])} \label{eq:upsa}
\end{align}
The learning algorithm of CoSam is presented in Algorithm \ref{al}.

\begin{algorithm}[t]
\small
\caption{Learning a recommendation model with CoSam}
\begin{algorithmic}[1]
\label{al}
\STATE Initialize parameters $\varphi,\theta$ randomly;
\WHILE {not converge}
\FOR{each user $u$}
\STATE Sample a set of items $G_u$ and paths $\mathcal S_u$ with adaptive random walk strategy.
\ENDFOR
\STATE Calculate gradients $\nabla_{\varphi}L$ w.r.t parameters $\varphi$ of R based on $G_u$ (eq. (\ref{eq:ups})).
\STATE Update $\varphi$ based on $\nabla_{\varphi}L$.
\STATE Calculate policy gradients $\nabla_{\theta}L$ w.r.t parameters $\theta$ of S based on $\mathcal S_u$ and $G_u$ (equation (\ref{eq:upsa})).
\STATE Update $\theta$ based on $\nabla_{\theta}L$.
\ENDWHILE
\end{algorithmic}
\end{algorithm}

\subsection{Discussion}
\label{se:dis}
\subsubsection{Gradient analysis.\\}
\begin{figure}[t]
  \centering
  \includegraphics[width=0.8\textwidth]{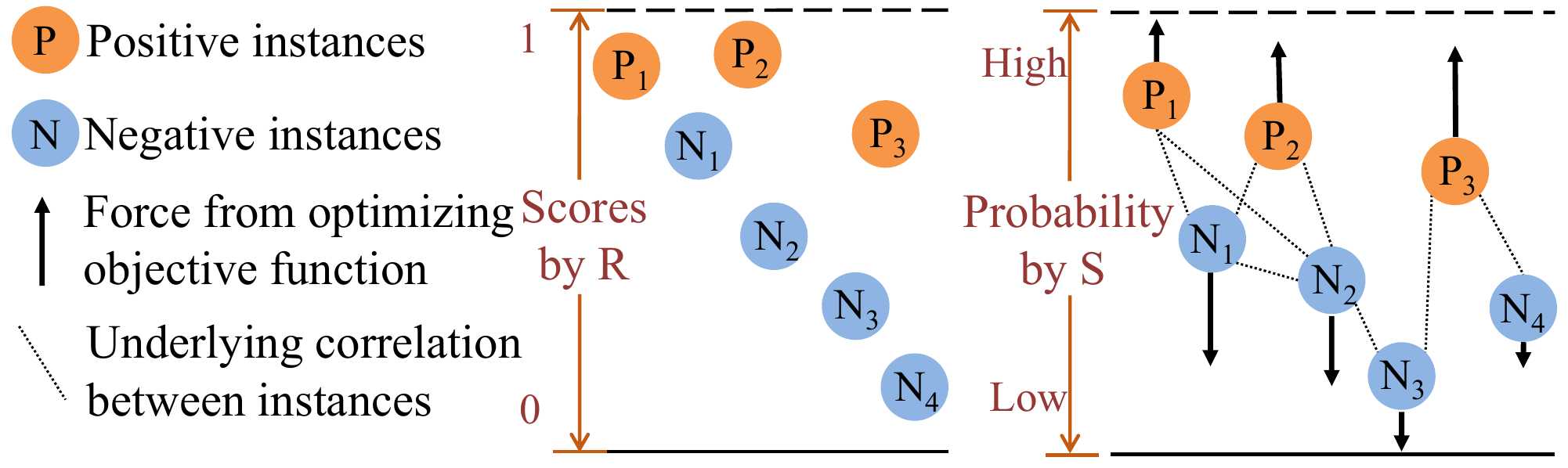}\\
  \caption{Illustration of how CoSam achieves sampling informatively and collaboratively.}\label{fg:an}
\end{figure}

\textbf{Sampling informatively.} How to understand the gradient of $L$ w.r.t parameters $\theta$ of S? Let us draw an analogy with the floating balls in the water as illustrated in Figure \ref{fg:an} for better description. The first term in eq.(\ref{eq:upr}) will give a force to push up these positive balls. Due to the correlations between the instances, which can be analogies with elastic links lines between the balls, the negative instances which have strong correlations with positive instances, will be pulled up due to the force from the lines. This effect drives S to over-sample "positive-like" instances, which can reduce the sampling variance and improve convergence.

\textbf{Sampling collaboratively.} On the other hand, from the second term in eq.(\ref{eq:upr}), the action of sampling a negative item will be punished with a negative reward $e_{ui}$. we observe an interesting phenomenon that when the sampled negative item is more hard to be differentiated by the model R, the reward will be more negative as illustrated in Figure \ref{fg:an}. It can be explained by the collaboration effect between S and R. Note that the negative items shall be differentiated either by the model S or by the model R. If R fails to differentiate a negative item, it naturally turn to S for help and gives a signal to decrease the probability of selecting this item into sampled item set. The reward guides S to flow to regions where S and R can well fit the data collaboratively. Moreover, if the sampling probability of the item decreases with the effect of the reward, it suggests the item can be differentiated by S and R's costly effort on this item will be spared; Otherwise, S will maintain relatively high sampling probability so that R can focus on this ``highly difficult'' instance.

\begin{figure}[t!]
  \centering
  \includegraphics[width=0.7\textwidth]{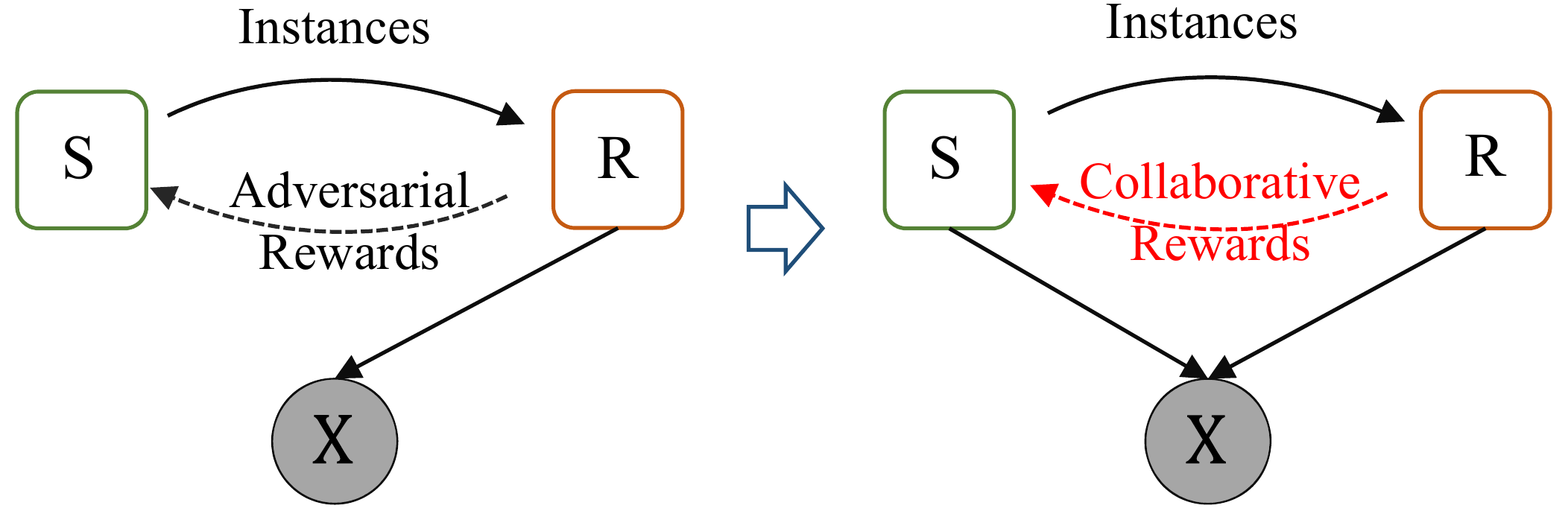}\\
  \caption{Illustration of the previous adversarial framework (left) and our integrated framework (right).}\label{fg:shif}
\end{figure}

\subsubsection{Complexity analysis.} The time cost on training CoSam consists of three parts: (1) On sampling, it requires $O(nl_r\bar N_u)$ time to sample instances and paths, where $\bar N_u$ denotes the average number of sampled instances over all users. (2) The model R can be updated with the sampled instances. This part depends on the selected recommendation model and its complexity is also linear to the number of sampled instances $O(n\bar N_u)$. (3) When updating the model S, we just calculate rewards and estimate policy gradients from the sampled instances. The time for this part is also $O(n\bar N_ul_r )$. In practise, as recent works \cite{chen2019samwalker,Yu2017}, we usually let $\bar N_u$ be several times (e.g. five) as large as the number of the positive items of the user $u$ and thus have $n\bar N_u=5|X^+|$, where $|X^+|$ denotes the number of observed positive data in the system. Overall, due to the sparsity of the recommendation data, our CoSam is efficient and can scale up to large dataset.

\subsubsection{Linking to existing methods. \\}

\textbf{Adversarial framework.}  We remark two key differences of our integrated framework from adversarial framework: (1) As illustrated in Figure \ref{fg:shif}, in our integrated framework, S and R collaboratively contribute to the prediction, while the prediction in adversarial framework just depends on R. (2) Both frameworks are trained in a reinforced manner. However, S in our framework is trained for collaboration with R; while S in adversarial framework is trained for competition with R.

\textbf{Two-stage recommendation strategy.} Another similar work is two-stage recommendation strategy \cite{asadi2012fast,covington2016deep,zhu2018learning,liu2019novel}. It contains: a recalling model to retrieve "positive-like" candidate items from global item pool and a re-ranking model to finely select recommendations from the candidate item set. This work is different from our integrated framework because the the two-stage recommendation strategy are not mechanism for training a recommendation model. Instead, two-stage recommendation strategy is a operation to accelerate retrieval from a pre-trained recommendation model by reducing the scale of the ranked items. Further, from the technical view, the intermediate item set in our framework is a ``soft'' probabilistic variable generated from the sampled model; while the intermediate item set in the two-stage recommendation strategy is a ``hard'' constant selected by the retrieval model.

\section{Related work}

\textbf{Recommendation from implicit feedback data.} Recent years have seen a surge of research on recommendation techniques from implicit feedback \cite{oard1998implicit,kelly2005implicit}. To handle the massive volume of the unobserved negative data, two strategies have been proposed in previous works. The first is employing memorization strategies to accelerate learning on the whole data. For example, ALS, eALS \cite{hu2008collaborative,he2016fast,bayer2017generic} have been proposed to memorize some specific intermediate variables to avoid the massive repeated computations. However, this kind of methods are just suitable for the recommendation model with K-separable property \cite{bayer2017generic} and L2 loss function. The other type is sampler-based methods, where the recommendation model are trained on the reduced sampling instances with stochastic gradient descent (SGD) optimizer. This kind of methods have been widely applied in many recommendation models, including classic matrix factorization \cite{hu2008collaborative,johnson2014logistic}, pair-wised models (e.g. Bpr  \cite{rendle2009bpr}, NCR\cite{song2018neural}) and sophisticated neural network-based methods (e.g. CDL \cite{wang2015collaborative}, DNN \cite{xue2017deep}, NCF \cite{he2017neural}, XDeepFM \cite{lian2018xdeepfm}, ConvNCF \cite{he2018outer}, NGCF \cite{wang2019neural},  lightGCN\cite{he2020lightgcn}, Mult-VAE \cite{liang2018variational}).

\textbf{Sampler for recommendation.} The most popular sampling strategy is to sample negative data with equal probability (uniformly). Although efficiently, the optimizer usually sample uninformative training instances \cite{rendle2014improving,yu2018walkranker,yu2017selection}, which makes limited contribution on updating. The model can not be well trained and the final recommendation performance will degrade. To deal with this problem, three types of sampling methods have been proposed in recent literatures. The first is heuristic-based strategies, which pre-define a sampling rule to over-sample informative instances \cite{yu2017selection,hernandez2014stochastic,yuan2017boostfm,zhao2014leveraging,rendle2014improving}. For example, \cite{yu2017selection,hernandez2014stochastic} propose to sample instances based on global item popularity; \cite{rendle2014improving, zhang2013optimizing, yu2018walkranker} propose to draw instances based on item ranking position. However, these methods requires rich human experience and also lack flexibility. Another type is model-based sampler \cite{wang2017irgan,park2019adversarial,ding2019reinforced}, which defines a sampler model and the sampling probability will adaptively evolve by optimizing adversarial objective. However, as analyzed in section 2, this kind of methods will suffer from inefficiency and biases. Also, there are some works leverage auxiliary information to sample informative training instances. For example, \cite{chen2019samwalker,wang2016social,zhao2014leveraging} claim to over-sample the items that have been consumed by social friends; \cite{ding2018improved,ding2019sampler,ding2019reinforced} leverage exposure data to find reliable negative instance. \cite{wang2020reinforced} leverages knowledge graph to enhance negative sampling. However, these auxiliary information may not be available in many recommendation systems.

\textbf{Random walk in recommendation.} Random walk strategy has been widely applied in recommendation. \cite{jamali2009trustwalker} performs random walk along the social network to search relevant users who have similar preference with the target user for better rating prediction;  \cite{christoffel2015blockbusters}  exploits random walk to obtain diverse recommendation; \cite{vahedian2017weighted} further extends \cite{christoffel2015blockbusters}  in heterogenous information network to generate valuable meta-paths; \cite{christoffel2015blockbusters} performs random walk to complete implicit feedback matrix for mitigating data sparsity problem; Similarly, \cite{yu2018walkranker} employs random walk to find more positive instances. We remark that these works adopt static (uniform or pre-defined) transfer probability in their random walk strategy. Besides, these random walks are not designed for sampling informative negative instances.

To our knowledge, only one recent work SamWalker \cite{chen2019samwalker} employs adaptive random walk to sample negative instances for better training of the recommendation model. The random walk in our CoSam differs from SamWalker in that: (1) the random walk in SamWalker is performed on the social network while it in CoSam is on the user-item interaction graph; (2) the transfer probability in SamWalker is learned from users' exposure while it in CoSam is learned with the rewards from the recommendation model.

\section{Experiments}

In this section, we conduct experiments to evaluate the performance of our CoSam. Our experiments are intended to address the following research questions:
\begin{enumerate}[(RQ1)]
\item  How does CoSam compare with existing sampling methods?
\item  Does CoSam boost recommendation performance?
\item How does different components in CoSam (i.e. collaborative sampling model and integrated framework) affect CoSam's performance?
\end{enumerate}

\subsection{Experimental settings}
\label{se:da}

\textbf{Datasets and parameters.} Four well-known recommendation datasets LastFM\footnote{\url{https://grouplens.org/datasets/hetrec-2011/ }}, Movielens-1M\footnote{\url{https://grouplens.org/datasets/movielens/}}, Ciao\footnote{\url{http://www.cse.msu.edu/~tangjili/trust}}, Epinions\footnote{\url{http://www.trustlet.org/epinions}}, are used in our experiments. These datasets contain users' feedback on items. The dataset statistics are presented in Table \ref{tb:da}. We preprocess the datasets so that all items have at least three interactions and "binarize" user's feedback into implicit feedback as \cite{xiao2017learning,chen2019samwalker}. That is, as long as there exists some user-item interactions (ratings or clicks), the corresponding implicit feedback is assigned a value of 1. Grid search and 5-fold cross validation are used to find the best hyper-parameters and test the results. In our CoSam, as recent works \cite{chen2019samwalker,Yu2017}, we set $N_u=5|\mathcal X_u|$ and test $c1,c2$ of [0.4,0.6,0.8,1]. We also optimize the model with mini-batch Adam \cite{kingma2014adam} and test the learning rate of [0.001,0.01,0.1]. The setting of compared methods are referring to related works or tested in our experiments. All experiments are conducted on a server with 2 Intel E5-2620 CPUs, 2 NVIDIA Titan-X GPUs and 256G RAM. We will share our source code online when the paper get published.

\begin{table}[t!]
\centering
\footnotesize
\caption{Statistics of four datasets.}

\label{tb:da}
\begin{tabular}{|c|c|c|c|c|}
\hline
Datasets                      & LastFM & Movielens & Ciao    & Epinions \\ \hline
Number of users               & 1,892  & 6,040        & 5,298   & 21,290   \\ \hline
Number of items               & 4,476  & 3,678       & 19,264  & 33,992   \\ \hline
Number of positive feedbacks  & 52,627 & 1,000,177    & 138,723 & 333,658  \\ \hline
Density of positive feedbacks & 0.62\% & 4.50\%       & 0.14\%  & 0.05\%   \\ \hline
\end{tabular}
\end{table}

\textbf{Evaluation Metrics.} We adopt the following metrics to evaluate recommendation performance:
\begin{itemize}
\item Recall@K (Rec@K): This metric quantifies the fraction of consumed items that are in the top-K ranking list sorted by their estimated rankings. For each user $i$, we define $Rec(i)$ as the set of recommended items in top-K and $Con(i)$ as the set of consumed items in test data for user $i$. Then we have:
\begin{align}
    Recall@K&=\frac{1}{{|U|}}\sum\limits_{i \in U} {\frac{{|Rec(i)\cap Con(i)|}}{|Con(i)|}}
\end{align}
\item Precision@K (Pre@K): This measures the fraction of the top-K items that are indeed consumed by the user:
 \begin{align}
    Precision@K&=\frac{1}{{|U|}}\sum\limits_{i \in U} {\frac{{|Rec(i)\cap Con(i)|}}{|Rec(i)|}}
\end{align}
\item Normalized Discounted Cumulative Gain (NDCG): This is widely used in information retrieval and it measures the quality of ranking through discounted importance based on positions. In recommendation, NDCG is computed as follow:
\begin{align}
NDCG &= \frac{1}{{|U|}}\sum\limits_{i \in U} {\frac{{DCG_i}}{{{IDCG_i}}}}
\end{align}
where ${DC{G_i}}$ is defined as follow and ${{IDCG_i}}$ is the ideal value of ${{DCG_i}}$ coming from the best ranking.
\begin{align}
{DCG_i} &= \sum\limits_{j \in  Con(i)} {\frac{1}{{{{\log }_2}(rank_{ij} + 1)}}}
\end{align}
where ${rank_{ij}}$ represents the rank of the item $j$ in the recommended list of the user $i$.
NDCG can be interpreted as the ease of finding all consumed items, as higher numbers indicate the consumed items are higher in the list.
\end{itemize}

\begin{table*}[t!]
\centering
\tiny
\caption{The performance metrics of the compared samplers. The boldface font denotes the winner in that column. The rows 'Impv' indicate the relative performance gain of CoSam over the best baseline. $\dagger$ indicates that the result of a paired difference test is significant at $p < 0.05$. We also present running time (second) for comparison.}

\label{tb:sa}
\begin{tabular}{|c|c|c|c|c|c|c|c|c|c|c|c|c|c|c|c|c|}
\hline
\multirow{2}{*}{Methods} & \multicolumn{4}{c|}{LastFM}                                                     & \multicolumn{4}{c|}{Movielens-1M}                                                          & \multicolumn{4}{c|}{Ciao}                                                       & \multicolumn{4}{c|}{Epinions}                                                   \\ \cline{2-17}
                         & Pre@5           & Rec@5           & NDCG            & \multicolumn{1}{l|}{Time} & Pre@5           & \multicolumn{1}{l|}{Rec@5} & NDCG            & \multicolumn{1}{l|}{Time} & Pre@5           & Rec@5           & NDCG            & \multicolumn{1}{l|}{Time} & Pre@5           & Rec@5           & NDCG            & \multicolumn{1}{l|}{Time} \\ \hline
Uniform                  & 0.0954          & 0.0859          & 0.3455          & 18.7                      & 0.3603          & 0.0790                     & 0.5669          & 280                       & 0.0140          & 0.0122          & 0.1775          & 65.0                      & 0.0085          & 0.0095          & 0.1557          & 156                       \\ \hline
Pop                      & 0.0978          & 0.0881          & 0.3513          & 21.6                      & 0.3633          & 0.0811                     & 0.5692          & 329                       & 0.0146          & 0.0119          & 0.1773          & 91.6                      & 0.0092          & 0.0098          & 0.1565          & 264                       \\ \hline
Adp                      & 0.0986          & 0.0881          & 0.3448          & 24.1                      & 0.3635          & 0.0802                     & 0.5667          & 315                       & 0.0146          & 0.0122          & 0.1784          & 94.5                      & 0.0088          & 0.0095          & 0.1559          & 348                       \\ \hline
Adv                      & 0.1002          & 0.0897          & 0.3503          & 104                       & 0.3691          & 0.0811                     & 0.5689          & 1212                      & 0.0153          & 0.0139          & 0.1786          & 980                       & 0.0094          & 0.0103          & 0.1567          & 8760                      \\ \hline
CoSam                    & \textbf{0.1302} & \textbf{0.1162} & \textbf{0.3838} & 23.2                      & \textbf{0.3772} & \textbf{0.0837}            & \textbf{0.5726} & 672                       & \textbf{0.0185} & \textbf{0.0149} & \textbf{0.1807} & 103                       & \textbf{0.0153} & \textbf{0.0168} & \textbf{0.1621} & 420                       \\ \hline
Impv\%                   & 23.00\%         & 29.53\%         & 9.24\%          & -                         & 2.17\%          & 3.15\%                     & 0.59\%          & -                         & 20.93\%         & 7.48\%          & 1.20\%          & -                         & 61.72\%         & 62.72\%         & 3.45\%          & -                         \\ \hline
\end{tabular}
\end{table*}

\subsection{Comparing with existing samplers (RQ1)}
To answer the research question (RQ1), we compare our CoSam with other sampling strategies including:
\begin{itemize}
\item Uniformly (Uni): the widely adopted sampling strategy which sample items with equal probability.
\item Item-Popularity (Pop) \cite{Yu2017}: which samples items based on item popularity: $p(i|u)\propto {\xi _i^\alpha }$, where $\xi _i$ denotes the popularity of item $i$ and the parameter $\alpha$ scales its affect.
\item Adaptive oversampling (Adp) \cite{rendle2014improving}: The Adaptive sampling strategy to over-sample the ``difficult'' items for the MF-based recommendation model.
\item Adversarial sampler (Adv): The state-of-the-art model-based sampler with adversarial sampler-recommender framework. Since the sampler model of \cite{ding2019reinforced} \cite{chen2019samwalker} requires other side information, we follow \cite{wang2017irgan} \cite{park2019adversarial} and model the sampler with matrix factorisation.
\end{itemize}
Also, for fairly comparison, we integrate these samplers into common matrix factorization recommendation model. Their performance are reported in Table \ref{tb:sa}.

\begin{figure*}[t]
  \centering
  \includegraphics[width=0.94\textwidth]{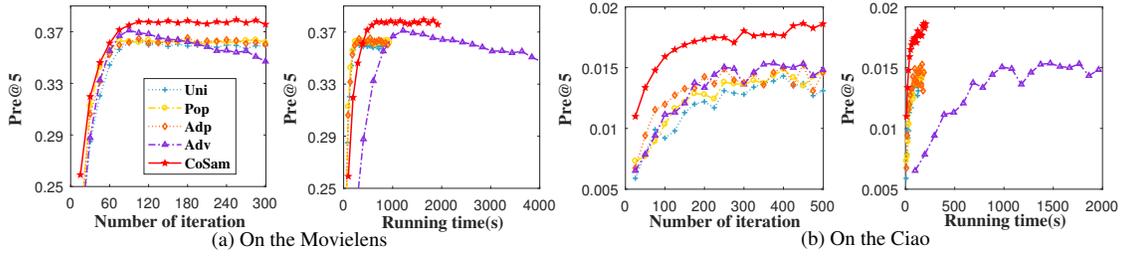}\\
  \caption{Pre@5 for different sampling strategy versus the number of iterations and running time.}\label{fg:sam}
\end{figure*}

\textbf{Performance comparison.} Table \ref{tb:sa} presents the recommendation performance of the compared sampler in terms of three evaluation metrics. The boldface font denotes the winner in that column. Overall, CoSam outperforms all compared methods on all datasets for all metrics. For the sake of clarity, the last row of Table \ref{tb:sa} also show the relative improvements achieved by CoSam over the best baselines. We have following observations: (1) The informative sampler will boost the recommendation performance. It can be seen from the better performance of Adv, Ada, CoSam over Uni. (2) The model-based samplers (CoSam and Adv), which will evolve with the training, outperform these heuristic methods. This result is consistent with our intuition, since the heuristic samplers lack flexible and usually fail to capture the varying informativeness of instances. (3) CoSam achieves much better performance than compared methods. Especially in the large dataset Epinions, the improvements is 61.7\%, 62.7\%, 3.45\% in terms of Pre@5, Rec@5, NDCG respectively. It can be attributed to collaborative sampler model and integrated sampler-recommender framework. We will conduct ablation studies in the subsection \ref{se:ab} to analyze the effect of these components.

\textbf{Running time.} Table \ref{tb:sa} also presents the training time cost of the compared samplers and figure \ref{fg:sam} depicts Pre@5 (Y-axis) vs. the number of iteration or running time (X-axis) of these methods on the two typically datasets. The powerful competitor Adv in accuracy spends much more time on sampling. Especially in the largest dataset Epinions, the recommendation method with Adv requires 2.4 hours for training while Cosam just requires 7 mins. 20 times acceleration will be achieved by our CoSam. The speed up can be attributed to our collaborative sampler, which avoids time-consuming traversal of all items. This way, CoSam has similar time complexity as these efficient heuristical methods, which aim at fast learning but sacrifice certain flexibility. Their actual running time are also in the same magnitude.

\begin{figure}[t]
  \centering
  \subfigure[Variance comparison]{
\includegraphics[width=0.425\textwidth]{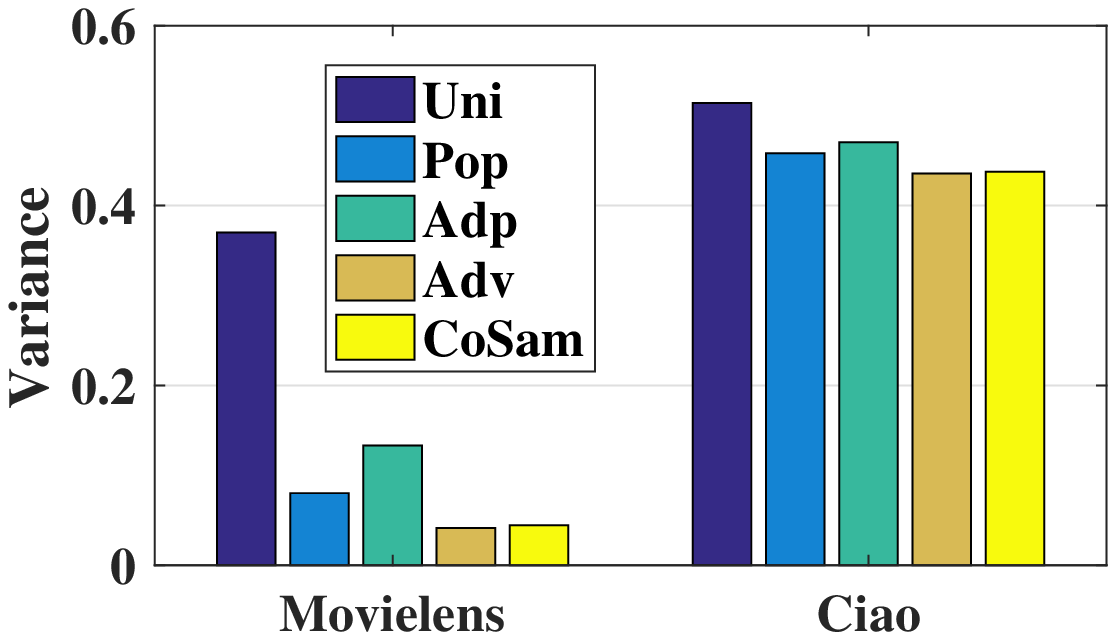}
}
\subfigure[Training loss comparison]{
\includegraphics[width=0.425\textwidth]{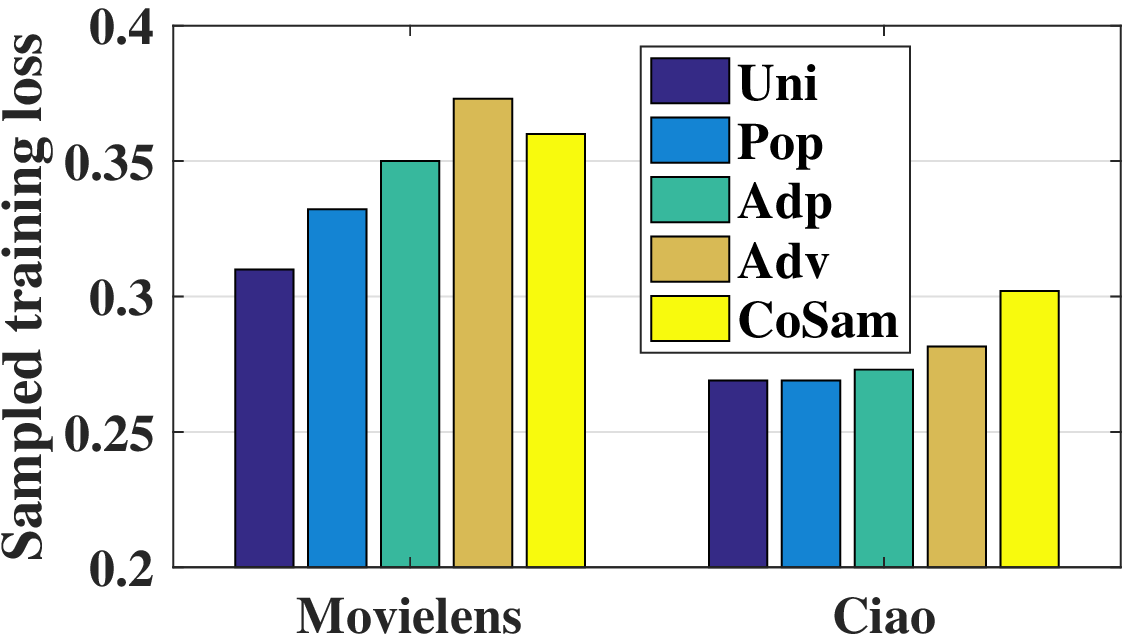}
}

\caption{Variance of gradients and sampled training loss using different sampling strategies.}\label{fg:va}
\end{figure}
\textbf{ Variance and training loss.} To show the stability and informativeness of samplers, we further conduct a specific experiments on empirically comparing the variance of the estimated gradient $\nabla _L\varphi$ and average training loss on sampled instances. To do so, we train the models for 50,100,150 epochs and then we generate mini-batch with various sampling strategy to estimate normalized gradients and training loss. We repeat this process for 1000 times, and calculate the variance of the gradients and average training loss for different sampling strategies. The final results are averaged over various parameters and presented in Figure \ref{fg:va}. The sampler value of variance suggests the model will be more stable, while the larger values of the sampled training loss indicates the sampled instances will be more informative. Figure \ref{fg:va} shows that our CoSam outperforms these heuristic samplers and achieves similar performance with Adv. We remark Adv will sacrifice training efficiency. Also, Adv will increase the risk of over-fitting as illustrated in Figure \ref{fg:sam}.

\subsection{Effect of CoSam (RQ2)}
To answer the research question (RQ2), we apply our CoSam sampling methods into two typical recommendation models (MF and NCF) and explore how CoSam boosts recommendation performance. We compare the improved recommendation methods MF-CoSam, NCF-CoSam with following recommendation methods:
\begin{itemize}
\item MF \cite{johnson2014logistic,hu2008collaborative,Pan2008}: The classic matrix factorization model for implicit feedback data with bernoulli likelihood and uniformly sampler.
\item BPR \cite{rendle2009bpr}: The classic pair-wise method for recommendation, coupled with matrix factorization.
\item NCF: \cite{he2017neural}: The advanced recommendation method modeling user-item interactions with neural network. We refer to \cite{he2017neural} and train the model with uniformly sampler.
\item IRGAN\cite{wang2017irgan}:  The method that employs generative adversarial model to capture users' preference over the items. Here we just report the results of IRGAN on the small dataset LastFM and Movielens due to its massive time cost on large datasets.
\item NGCF \cite{wang2019neural}: The state-of-the-art graph-based recommendation method, which encodes users and items based on neural graph network \cite{hamilton2017inductive} on the user-item interaction graph.
\end{itemize}

\begin{figure}[t]
  \centering
  \subfigure{
\includegraphics[width=0.9\textwidth]{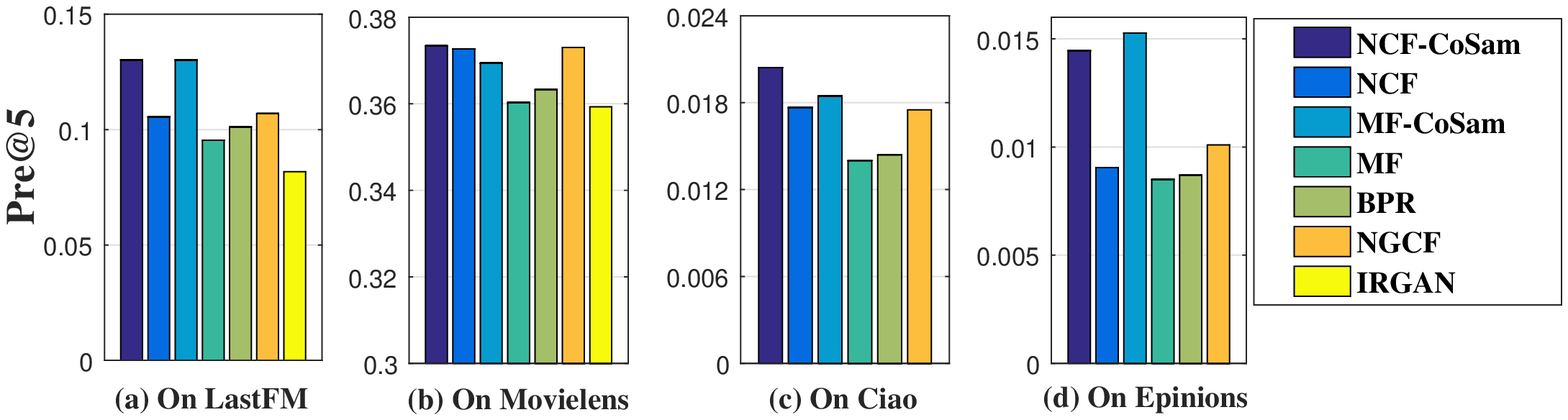}
}
\subfigure{
\includegraphics[width=0.9\textwidth]{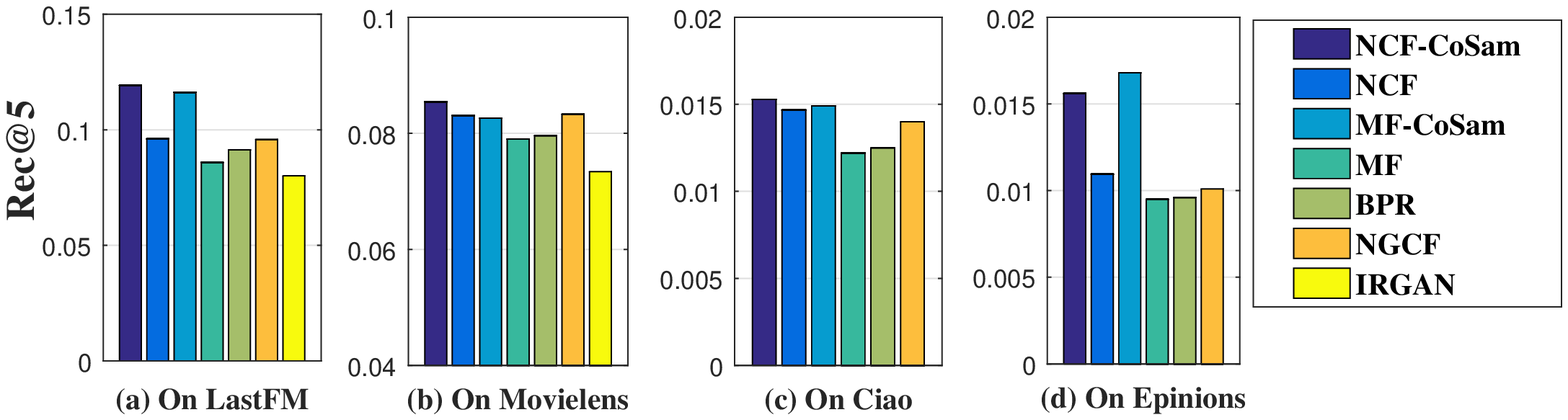}
}
  \caption{Performance of different recommendation methods.}\label{fg:rs}
\end{figure}

Figure \ref{fg:rs} presents the recommendation performance of the compared methods. The improvement from CoSam is apparent: adopting CoSam consistently boost the recommendation performance. Especially on the large dataset Epinions, the improvements are over 30\% achieved by MF-CoSam (or NCF-CoSam) compared with MF (or NCF). Further, we observe that these sampler-enhanced method can beat state-of-the-art recommendation methods, even if we just employ a relatively simple recommendation model (i.e. MF). This result validate the importance of the sampling method, which makes nonnegligible contribution on recommendation performance. In fact, the recommendation model has been well exploited by substantial recent works while the sampler usually become the effect bottleneck of the recommendation system.

\begin{table*}[t!]
\centering
\tiny
\caption{The performance and running time of CoSam and its variants.}
\label{tb:sp}
\begin{tabular}{|c|c|c|c|c|c|c|c|c|c|c|c|c|c|c|c|c|}
\hline
\multirow{2}{*}{Methods}                                    & \multicolumn{4}{c|}{LastFM}                          & \multicolumn{4}{c|}{Movielens-1M}                                        & \multicolumn{4}{c|}{Ciao}                            & \multicolumn{4}{c|}{Epinions}                        \\ \cline{2-17}
                                                            & Pre@5  & Rec@5  & NDCG   & \multicolumn{1}{l|}{Time} & Pre@5  & \multicolumn{1}{l|}{Rec@5} & NDCG   & \multicolumn{1}{l|}{Time} & Pre@5  & Rec@5  & NDCG   & \multicolumn{1}{l|}{Time} & Pre@5  & Rec@5  & NDCG   & \multicolumn{1}{l|}{Time} \\ \hline
\begin{tabular}[c]{@{}c@{}}Sam-MF-Adv\\ (Adv)\end{tabular}  & 0.1002 & 0.0897 & 0.3503 & 104                       & 0.3691 & 0.0811                     & 0.5689 & 1414                      & 0.0153 & 0.0139 & 0.1786 & 980                       & 0.0094 & 0.0103 & 0.1567 & 8760                      \\ \hline
Sam-MF-Int                                                   & 0.1073 & 0.0957 & 0.3579 & 116                       & 0.4043 & 0.0953                     & 0.5962 & 1484                      & 0.0160 & 0.0131 & 0.1797 & 977                       & 0.0125 & 0.0134 & 0.1545 & 9984                      \\ \hline
Sam-Co-Adv                                                  & 0.0994 & 0.0892 & 0.3454 & 19                        & 0.3643 & 0.0805                     & 0.5676 & 630                       & 0.0154 & 0.0138 & 0.1788 & 102                       & 0.0091 & 0.0098 & 0.1562 & 322                       \\ \hline
\begin{tabular}[c]{@{}c@{}}Sam-Co-Int\\ (CoSam)\end{tabular} & 0.1302 & 0.1162 & 0.3838 & 23                        & 0.3772 & 0.0837                     & 0.5726 & 672                       & 0.0185 & 0.0149 & 0.1807 & 103                       & 0.0153 & 0.0168 & 0.1621 & 420                       \\ \hline
\end{tabular}
\end{table*}

\begin{table}[t!]
\centering
\footnotesize
\caption{Characteristics of Sam-MF-Int and its variants.}
\label{tb:ch}
\begin{tabular}{|c|c|c|c|}
\hline
Methods      & Training criterion & \begin{tabular}[c]{@{}c@{}}Prediction\\ with S?\end{tabular} & \begin{tabular}[c]{@{}c@{}}Prediction\\ with R?\end{tabular} \\ \hline
Sam-MF-Int    & Integrated      & ${\surd}$                                                    & ${\surd}$                                                    \\ \hline
Sam-MF-Int-S  & Integrated       & ${\surd}$                                                    & $\backslash$                                                 \\ \hline
Sam-MF-Int-R  & Integrated       & $\backslash$                                                 & ${\surd}$                                                    \\ \hline
Sam-MF-adv-U & Adversarial        & ${\surd}$                                                    & ${\surd}$                                                    \\ \hline
Sam-MF-adv   & Adversarial        & $\backslash$                                                 & ${\surd}$                                                    \\ \hline
\end{tabular}
\end{table}

\subsection{Ablation study (RQ3)}
\label{se:ab}
 We further design a detailed ablation study to answer question RQ3. That is, we remove different components at a time and compare CoSam with its three special cases: Sam-MF-Int, Sam-Co-Adv, Sam-MF-Adv. Here we remark Sam-X-Y as the sampler with ``X'' sampler model and ``Y'' sampler-recommender framework. Here ``MF'' denotes matrix factorization model and ``Co'' denotes collaborative model; ``Adv'' denotes Adversarial framework and ``Int'' denotes integrated framework. Note that Cosam can be remarked as Sam-Co-Int and Adv can be remarked as Sam-MF-Adv. We integrate these methods into benchmark MF recommendation model to test their performance. The recommendation performance and training running time are presented in Table \ref{tb:sp}.

We observe that the method with integrated sampler-recommender framework (Sam-Co-Int, Sam-MF-Int) outperforms their variants with adversarial framework (Sam-Co-Adv, Sam-MF-Adv). This is due to the ability of the sampler model has been better exploited and the sampling biases have been refined in the collaborative framework. For the sampler models comparison, generally, the collaborative sampler model outperforms MF-based sampler in the collaborative framework, while it performs worse in the adversarial framework. This interesting phenomenon can be explained as follow: collaborative sampler, which encodes rich interaction information into sampling, will produce more informative instances. Naturally, it will achieves better performance under integrated framework. However, when it is applied into adversarial framework, this informative but highly biased sampler will heavily deteriorate the performance and offset the advantage from informativeness. It is noteworthy that the collaborative sampler model performs much more efficient than MF-based sampler model. Thus, collaborative sampler will be more suitable for the modern recommendation system with large item space.

\textbf{Effect of integrated recommendation.} Another interesting ablation experiments has been conducted to show the effect of integrating S and R into prediction for recommendation. That is, we compare Sam-MF-Int with following variants:
\begin{itemize}
\item Sam-MF-Int-S, Sam-MF-Int-R: The special cases of Sam-MF-Int, which trains S and R based on integrated framework but make recommendation just based on S (Sam-MF-Int-S) or R (Sam-MF-Int-R).
\item Sam-MF-Adv-U: The variant of Sam-MF-Adv, which trains S and R based on adversarial framework but unifies S and R for recommendation as integrated framework.
\end{itemize}
Their characteristics are presented in Table \ref{tb:ch}. Here we choose Sam-MF-Int instead of Sam-Co-Int (CoSam) for experiments to isolate the effect of collaborative sampler model.

\begin{figure}[t]
  \centering
  \includegraphics[width=0.8\textwidth]{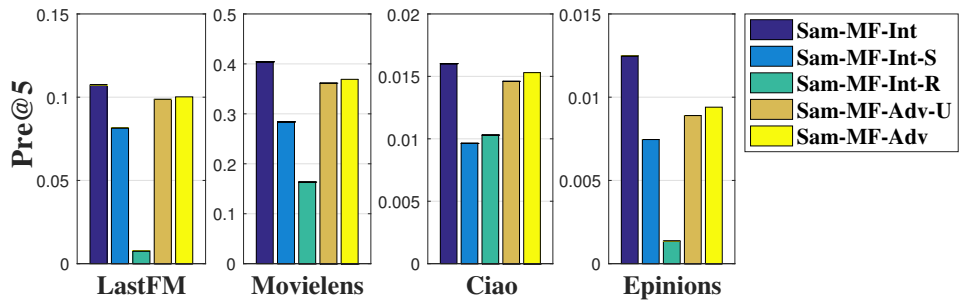}\\
  \caption{Performance of Sam-MF-Int and its variants.}\label{fg:sp}
\end{figure}

Figure \ref{fg:sp} presents the performance of these compared methods. We observe that Sam-MF-Int consistently outperforms Sam-MF-Int-R and Sam-MF-Int-S, which trains S and R collaboratively but make recommendations separately. This result validates the effectiveness of unifying S and R into prediction. The performance will degrade if we only consider one part. R will be biased to focus on these ``difficult'' instances and its performance on other relatively ``easy'' will be deteriorated; Similarly, S is not trained to capture users' preference individually and thus Sam-MF-Int-S performs inferior to the many recommendation method. Only when R meets S in recommendation, they will compensate each other and achieve impressive performance. However, we observe that unifying R and S into recommendation with adversarial framework is ineffective. Sam-MF-adv-U performs worse instead of better than Sam-MF-adv. It can be attributed to the different rewards (as discussed in subsection \ref{se:dis}). In adversarial framework, S is trained for competition instead of collaboration with R. S and R are not ready for collaborative prediction.

\section{Conclusion}
In this paper, we present a novel efficient and effective sampling method CoSam. It consists of: (1) a collaborative sampler model which leverages collaborative signals into sampling and satisfies the desirable property of normalization, adaption, interaction information awareness, and computational efficiency; (2) an integrated sampler-recommender framework to refine the sampling bias and to better exploit the ability of the sampler model. Our extensive experiments on four real-world datasets validate the superiority of the proposed collaborative sampler and integrated framework.

One interesting direction for future work is to leverage sophisticated graph neural network in sampler process, to capture more complex affinity between users and items along the interaction graph. Also, effectiveness and efficiency are dual objectives for sampling strategies. It will be interesting and challenging to design more advanced sampler meeting both conditions. Another interesting direction is to explore dynamic sampling strategy in online recommender system, where user preferences will drift with time\cite{lei2020estimation}.

\section*{ACKNOWLEDGMENTS}
This work is supported by National Key R\&D Program of China (Grant No: 2018AAA0101505, 2019YFB1600700) and National Natural Science Foundation of China (Grant No: U1866602).

\bibliographystyle{ACM-Reference-Format}
\bibliography{sigproc}
\end{document}